\documentclass[10pt,twocolumn,twoside]{IEEEtran} 
\ifCLASSINFOpdf
   \usepackage[pdftex]{graphicx}
\else
   \usepackage[dvips]{graphicx}
\fi
\usepackage[cmex10]{amsmath}
\usepackage{amsfonts}
\newcommand{\vc}[1]{{\boldsymbol{#1}}}
\newcommand{\re}{{\mathbb{R}}}
\newcommand{\D}{\mathcal{D}}
\newcommand{\dd}{{\mathrm{d}}}
\newcommand{\ipr}[2]{\langle #1, #2 \rangle}

\newtheorem{problem}{Problem}
\newtheorem{lemma}{Lemma}
\newtheorem{remark}{Remark}

\begin{document}
%
\title{$L^1$ Control Theoretic Smoothing Splines}
%
%
%

\author{Masaaki~Nagahara,~\IEEEmembership{Member,~IEEE,}
        and~Clyde~F.~Martin,~\IEEEmembership{Fellow,~IEEE}
\thanks{Copyright (c) 2012 IEEE. Personal use of this material is permitted. However, permission to use this material for any other purposes must be obtained from the IEEE by sending a request to pubs-permissions@ieee.org.}        
\thanks{M. Nagahara is with
	Graduate School of Informatics,
	Kyoto University, Kyoto, 606-8501, Japan;
	email: nagahara@ieee.org (corresponding author)}
\thanks{C. F. Martin is with
Department of Mathematics \& Statistics, Texas Tech University,
Texas, USA; email: clyde.f.martin@ttu.edu}
}

%
%

\markboth{Journal of \LaTeX\ Class Files,~Vol.~11, No.~4, December~2012}%
{Shell \MakeLowercase{\textit{et al.}}: Bare Demo of IEEEtran.cls for Journals}
%



\maketitle

\begin{abstract}
In this paper, we propose control theoretic smoothing splines 
with $L^1$ optimality for reducing the number of parameters
that describes the fitted curve as well as removing outlier data.
A control theoretic spline is a smoothing spline
that is generated as an output of a given linear dynamical system.
Conventional design requires exactly the same number of
base functions as given data, and the result is not robust
against outliers.
To solve these problems,
we propose to use $L^1$ optimality, that is,
we use the $L^1$ norm for the regularization term
and/or the empirical risk term.
The optimization is described by a convex optimization,
which can be efficiently solved via a numerical optimization software.
A numerical example shows the effectiveness of the proposed method.
\end{abstract}


%
\IEEEpeerreviewmaketitle

\section{Introduction}
%
%
%
%

The spline has been widely used in signal processing,
numerical computation, statistics, etc.
In particular,
the {\em smoothing spline} gives a smooth curve
that has the best fit to given noisy data
\cite{KimWah71,Wah}.
The smoothness is achieved by
limiting the $L^2$ norm of the $m$-th
derivative of the curve as well as minimizing
the squared error (or empirical risk) between data and the curve.

\begin{figure}[t]
 \centering
 \includegraphics[width=0.6\linewidth]{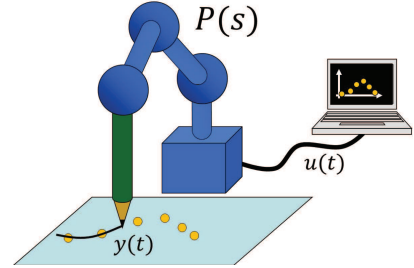}
 \caption{Control theoretic spline as a robot $P(s)$ that draws a smooth curve $y(t)$
 with a control input $u(t)$ based on given data.}
 \label{fig:cts}
\end{figure}

The {\em control theoretic smoothing spline} \cite{SunEgeMar00}
is generalization
of the smoothing spline
using control theoretic ideas,
by which the spline curve is determined by
the output of a linear dynamical system.
It is shown in \cite{EgeMar} that control theoretic splines give a richer class of smoothing curves
relative to polynomial curves.
Fig.~\ref{fig:cts} illustrates the idea of the control theoretic spline;
given a finite number of data,
the robot modeled by a dynamical system with transfer function $P(s)$ is driven by
a control input $u(t)$ and draws a smooth curve $y(t)$
that fits to the data.
The problem of the control theoretic spline is to find control $u(t)$
that gives an expected motion of the robot,
based on the model $P(s)$ and the data set.
Furthermore, 
the control theoretic spline has been proved to be useful for trajectory planning in \cite{EgeMar01},
mobile robots in \cite{TakMar04}, 
contour modeling of images in \cite{KanEgeFujTakMar08},
probability distribution estimation in \cite{Cha10}, to name a few. 
For more
applications and a rather complete theory of control theoretic
splines, see \cite{EgeMar}.

Conventional design of control theoretic splines is based on $L^2$ optimization
\cite{SunEgeMar00}, and has two main drawbacks.
One is that we need the same number
of parameters as the data
to represent the fitted curve.
If the data set is big, then the number of parameters
becomes crucial when for example the actuator system of the robot (see Fig.~\ref{fig:cts})
has just a small area of memory.
The other drawback is that the spline is not robust against outliers
in observed data. In other words, conventional control theoretic splines are
sensitive to outliers.
To overcome these drawbacks,
we propose to use $L^1$ optimality in the design.
For reduction of the number of parameters,
we utilize the {\em sparsity-promoting property}
of the $L^1$ norm regularization,
also known as LASSO (least absolute shrinkage and selection operator)
\cite{Tib96,BuhGee}.
For robustness against outliers,
we adopt the $L^1$ norm for the empirical risk minimization
\cite{SchSmo},
assuming that the noise is Laplacian, 
heavier-tailed distribution than Gaussian that is assumed in conventional studies%
\footnote{The idea of using a heavier-tailed loss function
for control theoretic smoothing splines was first
proposed in \cite{Nag11,NagMar13WSC}.}.
The problem is then described in convex optimization,
which can be efficiently solved by numerical computation software,
e.g. \texttt{cvx} on MATLAB \cite{cvx,GraBoy08}.
For numerical computation, we implement the design procedure on
MATLAB programs with \texttt{cvx},
access \cite{matlab} to obtain the programs.
Based on the programs, we show a numerical example 
that illustrates the effectiveness of the proposed method.

The remainder of this article is
organized as follows:
Section~\ref{sec:L2} reviews the conventional $L^2$-optimal control theoretic spline
and discusses drawbacks of the $L^2$ spline.
Section~\ref{sec:L1} formulates the problem of the proposed $L^1$ spline
to overcome drawbacks in the $L^2$ spline,
and show a procedure to the solution.
A numerical example is included in 
Section \ref{sec:simulation}. 
Section \ref{sec:conclusion} draws conclusions.

\section{$L^2$ Control Theoretic Smoothing Splines}
\label{sec:L2}

Consider a linear dynamical system $P$ defined by
\begin{equation}
 \begin{split}
  \dot{\vc{x}}(t) &= A\vc{x}(t) + \vc{b}u(t),\quad \vc{x}(0)=\vc{0}\in\re^{n},\\
  y(t) &= \vc{c}^\top \vc{x}(t),  \quad t\geq 0
   \end{split}
 \label{eq:system}
\end{equation}
where $A\in\re^{n\times n}$, $\vc{b},\vc{c}\in\re^{n}$.
We assume $(A,\vc{b})$ is controllable and $(\vc{c}^\top,A)$ is observable%
\footnote{For controllability and observability of a linear system,
see e.g.\ \cite[Chap.~9]{Rug}.}.
For this system, suppose that a data set
\[
 \D := \{(t_1, y_1), (t_2, y_1), \ldots (t_N, y_N)\}
\] 
is given,
where $t_1,\ldots,t_N$ are sampling instants which satisfy
$0<t_1<t_2<\cdots<t_N=:T$,
and $y_1,\ldots,y_N$ are noisy sampled data
of the output of \eqref{eq:system}.
The objective here is to find control $u(t)$, $t\in[0,T]$ for the dynamical
system \eqref{eq:system}
such that
$y(t_i)\approx y_i$ for $i=1,\ldots,N$.
For this purpose, the following \emph{quadratic} cost function has been introduced in 
\cite{SunEgeMar00}:
\begin{equation}
 J(u) := \lambda \int_0^T |u(t)|^2 \dd t + \sum_{i=1}^N w_i |y(t_i) - y_i|^2,
 \label{eq:J2}
\end{equation}
where 
$\lambda>0$ is the regularization parameter that specifies the tradeoff between
the smoothness of control $u(t)$ defined in the first term of \eqref{eq:J2}
and the minimization of the squared empirical risk in the second term.
Also, $w_i>0$ is a weight for $i$-th squared loss $|y(t_i)-y_i|^2$.
Then the problem of $L^2$ control theoretic smoothing spline
is formulated as follows:
\begin{problem}[$L^2$ control theoretic smoothing spline]
Find control $u(t)$ that minimizes the cost $J(u)$ in \eqref{eq:J2} subject to
the state-space equation in \eqref{eq:system}.
\end{problem}

The optimal control $u=u^\ast$ that minimizes $J(u)$ is given by
\cite{SunEgeMar00,EgeMar}
\begin{equation}
 u^\ast(t) = \sum_{i=1}^N \theta^\ast_i g(t_i-t),
 \label{eq:optimal_u2}
\end{equation}
where $g(\cdot)$ is defined by
\begin{equation}
 g(\tau) := \begin{cases} 
	\vc{c}^\top e^{A\tau}\vc{b},
	&\quad \text{if}~~\tau\in[0,T],\\ 
	0,&\quad \text{otherwise.}
 \end{cases}
 \label{eq:gi}
\end{equation}
Note that $\vc{c}^\top e^{A\tau} \vc{b}$ in $g(\tau)$ is the impulse response
of the dynamical system \eqref{eq:system}.
The optimal coefficients $\theta_1^\ast,\dots,\theta_N^\ast$ are given by
\begin{equation}
 \vc{\theta}^\ast:=\bigl[\theta^\ast_1,\ldots,\theta^\ast_N\bigr]^\top=(\lambda I + WG)^{-1}W\vc{y},
 \label{eq:L2optimal}
\end{equation}
where
\begin{equation}
W:={\mathrm{diag}}(w_1,\ldots,w_N),
\quad \vc{y}:=[y_1,\ldots, y_N]^\top.
\label{eq:y_vec}
\end{equation}
The matrix $G=[G_{ij}]\in \re^{N\times N}$ in \eqref{eq:L2optimal}
is the Grammian defined by 
\begin{equation}
 \begin{split}
 G_{ij}&=\ipr{g(t_i-\cdot)}{g(t_j-\cdot)}\\
  &=\int_0^T g(t_i-t)g(t_j-t)\dd t,\quad i,j=1,\ldots,N.
 \end{split} 
 \label{eq:Gij}
\end{equation}

An advantage of the $L^2$ control theoretic smoothing spline is that 
the optimal control can be computed offline via equation \eqref{eq:L2optimal}.
However, the formula indicates that if the data size $N$ is large,
so is the number of base functions in $u^\ast(t)$, 
as shown in \eqref{eq:optimal_u2}.
This becomes a drawback if we have only a small memory or a
simple actuator for drawing a curve with the optimal control $u^\ast(t)$.
Another drawback is that the $L^2$ spline is not robust at all
against outliers, as reported in \cite{NagMar13WSC},
since the squared empirical risk in \eqref{eq:J2} assumes
that the additive noise is Gaussian.
To solve these problems, we adopt $L^1$ optimality
for the design of spline.

\section{$L^1$ Control Theoretic Smoothing Splines}
\label{sec:L1}
Before formulating the design problem of $L^1$ spline, we prove the following lemma:
\begin{lemma}
Assume that control $u(t)$ is given by
\begin{equation}
 u(t) = \sum_{i=1}^N \theta_i g(t_i-t),
  \label{eq:u}
\end{equation} 
for some $\theta_i\in\re$, $i=1,2,\ldots,N$.
Then we have
\begin{equation}
 y(t) = \sum_{i=1}^N \theta_i\ipr{g(t-\cdot)}{g(t_i-\cdot)},\quad t \in [0,T].
 \label{eq:y}
\end{equation}
In particular, for $j=1,2,\ldots,N$, we have
\begin{equation}
 y(t_j) = \sum_{i=1}^N \theta_iG_{ij}.
 \label{eq:yti}
\end{equation}
\end{lemma}
\begin{IEEEproof}
If $u(t)=0$ for $t<0$, then 
the solution of \eqref{eq:system} is given by
\[
 \begin{split}
 y(t) = \int_0^t \vc{c}^\top e^{A(t-\tau)}\vc{b}u(\tau) \dd \tau
 &= \int_0^T g(t-\tau)u(\tau)\dd \tau\\
 &= \ipr{g(t-\cdot)}{u}
 \end{split}
\] 
Substituting \eqref{eq:u} into the above equation gives
\eqref{eq:y}.
Then, from the definition of $G_{ij}$ in \eqref{eq:Gij},
we immediately have \eqref{eq:yti}.
\end{IEEEproof}

By this lemma, the error $y(t_i)-y_i$ is given by
\[
 y(t_j)-y_j = \sum_{i=1}^N\theta_iG_{ij}-y_i,\quad j=1,2,\ldots,N,
\]
or equivalently
\begin{equation}
 \begin{bmatrix}y(t_1)-y_1\\\vdots\\y(t_N)-y_N\end{bmatrix} = G\vc{\theta}-\vc{y},
 \label{eq:vector_error}
\end{equation}
where $\vc{\theta}:=[\theta_1,\ldots,\theta_N]^\top$ and $\vc{y}$ is given in \eqref{eq:y_vec}.
Based on this, we consider the following optimization problem:
\begin{problem}[$L^1$-optimal spline coefficients]
\label{prob:L1}
Find $\vc{\theta}\in\re^N$ that minimizes
\begin{equation}
 J_{p}(\vc{\theta}) :=\eta\|\vc{\theta}\|_1 + \|W(G\vc{\theta}-\vc{y})\|_p^p,
 \label{eq:Jp}
\end{equation}
where $\eta>0$ and $p\in\{1,2\}$.
\end{problem}

The regularization term, $\|\vc{\theta}\|_1$,
is for sparsity of coefficients $\theta_1,\ldots,\theta_N$,
as used in LASSO \cite{Tib96,BuhGee}.
Also, small $\|\vc{\theta}\|_1$ leads to small $L^1$ norm
of control $u$ since from \eqref{eq:u} we have
\[
 \int_0^T |u(t)|\dd t \leq C \|\vc{\theta}\|_1,
\]
for some constant $C>0$.
On the other hand, the empirical risk term,
$\|W(G\vc{\theta}-\vc{y})\|_p^p$,
is for the fidelity to the data.
For $p=1$, additive noise is assumed to be Laplacian,
a heavy-tailed distribution,
to take outliers into account,
while $p=2$ is related to Gaussian noise.
In each case, cost function $J_p(\vc{\theta})$
is convex in $\vc{\theta}$.

Unlike $L^2$ spline, the solution to the optimization in Problem \ref{prob:L1} cannot be
represented in a closed form.
However, by using a numerical optimization algorithm we can obtain 
an approximated solution within a reasonable time.
For example, for the optimization with $p=2$, we can adopt
FISTA (Fast Iterative Shrinkage-Thresholding Algorithm)
\cite{BecTeb09},
which is an extension of Nesterov's work \cite{Nes83}
to achieve the convergence rate $O(1/k^2)$ at $k$-th iteration.
On the other hand, for $p=1$, 
there is no algorithm achieving such a rate, but
the optimization is still convex
and we can use an efficient convex optimization software, such as
\texttt{cvx} on MATLAB \cite{cvx,GraBoy08}.

\begin{remark}
The optimization is related to the following signal subspace
\[
 V := \biggl\{u\in L^2[0,T]: u=\sum_{i=1}^N \theta_ig(t_i-\cdot),~ \theta_i\in\re\biggr\}.
\]
That is, we seek the optimal control $u$ in $V$ such that 
the coefficients minimize \eqref{eq:Jp}.
Note that $\{g(t_1-\cdot),\ldots,g(t_N-\cdot)\}$
is a basis of $V$ due to the controllability and observability of
system \eqref{eq:system}.
\end{remark}

\begin{remark}
Although we have assumed that the initial state $\vc{x}$ is $\vc{0}$,
we can also set the initial state $\vc{x}(0)=\vc{x}_0$ 
as a design variable in a similar manner.
In this case, the output $y(t)$ becomes
\[
 y(t_j) = \vc{c}^\top e^{At_j} \vc{x}_0 + \sum_{i=1}^N \theta_iG_{ij},\quad j=1,2,\ldots,N,
\]
and the optimization is formulated by
\begin{equation}
 \min_{\vc{x}_0,\vc{\theta}}\left\{ \eta\|\vc{\theta}\|_1 + \|W(H\vc{x}_0+G\vc{\theta}-\vc{y})\|_p^p\right\},
 \label{eq:opt_initial}
\end{equation}
where $H:=[e^{A^\top t_1}\vc{c},\ldots,e^{A^\top t_N}\vc{c}]^\top$.
This is also a convex optimization problem and can be efficiently solved via numerical optimization softwares.
\end{remark}
\begin{remark}
The choice of parameters $\eta$ and $w_i$ influences
the performance of curve fitting.
The regularisation parameter $\eta$ controls the trade-off
between the sparsity and fidelity of the solution;
a larger $\eta$ leads to a sparser solution (i.e. more $\theta_i$'s are zero)
while a smaller $\eta$ leads to a smaller empirical risk.
On the other hand, $w_i$ may be chosen to be larger
if the data $y_i$ contains smaller error.
These parameters should be chosen by trial and error
(e.g. cross-validation \cite{BuhGee}).
\end{remark}

\section{Numerical Example}
\label{sec:simulation}
\begin{figure}[t]
\centering
\includegraphics[width=\linewidth]{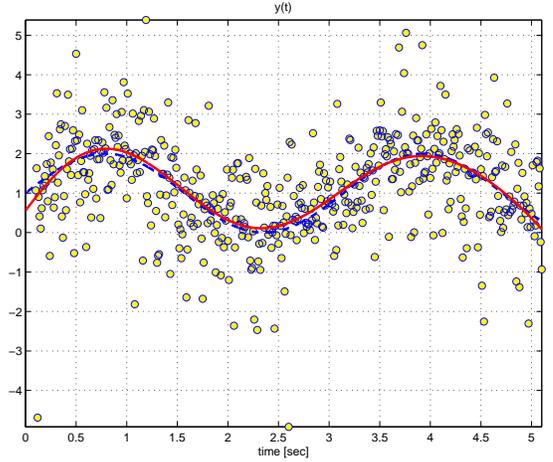}
\caption{Simulation result of $L^1$ spline: original curve (dashed line), observed data (circles), fitted curve (solid line).}
\label{fig:L1result}
\end{figure}

In this section, we show a numerical example that
illustrates the effectiveness of the proposed $L^1$ control theoretic
smoothing spline.
We set the dynamical system $P$ with transfer function
\[
 P(s) = \frac{1}{s^3+1}.
\]
State-space matrices for $P(s)$ are given by
\[
 A = \begin{bmatrix}0 & 0 & -1\\ 1 & 0 & 0\\ 0 & 1 & 0 \end{bmatrix},\quad
 \vc{b} = \begin{bmatrix} 1 \\ 0 \\ 0 \end{bmatrix},\quad
 \vc{c} = \begin{bmatrix} 0 \\ 0 \\ 1 \end{bmatrix}.
\]
We assume the original curve is given by
\[
 y_{{\mathrm{orig}}}(t) = \sin(2t) + 1.
\]
The sampling instants are given by
\[
 t_i = 0.1 + 0.01(i-1), \quad i=1,2,\ldots,501,
\]
that is, 
the data are sampled at rate $100$ [Hz] (100 samples per second) 
from initial time $t_1=0.1$.
The observed data $y_1,y_2,\ldots,y_{501}$ are assumed to be
disturbed by additive Laplacian noise with mean $0$ and variance $1$.
See Fig.~\ref{fig:L1result} for the original curve $y_{{\mathrm{orig}}}(t)$
and the observed data $y_1,y_2,\ldots,y_{501}$.

For these data, we compute the optimal coefficients of the $L^1$ control theoretic spline
with $p=1$ corresponding to Laplacian noise.
The design parameters are $\eta=0.01$ and $w_i=1$ for all $i$
(i.e. all elements have equal weight).
We assume that the initial state $\vc{x}(0)=\vc{x}_0$ is also a
design variable, that is, we solve optimization \eqref{eq:opt_initial}.

\begin{figure}[t]
\centering
\includegraphics[width=\linewidth]{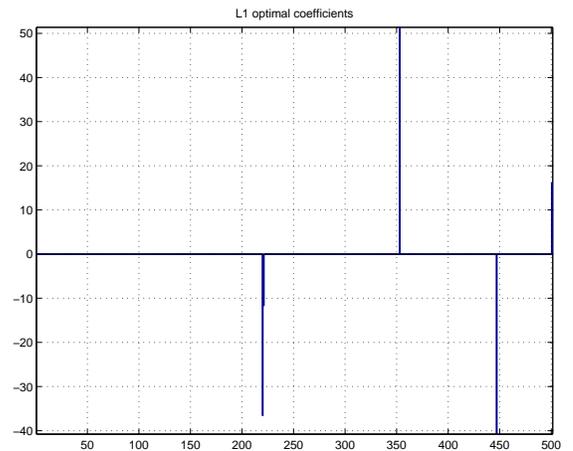}
\caption{Coefficients of $L^1$ spline}
\label{fig:L1coef}
\end{figure}

\begin{figure}[t]
\centering
\includegraphics[width=\linewidth]{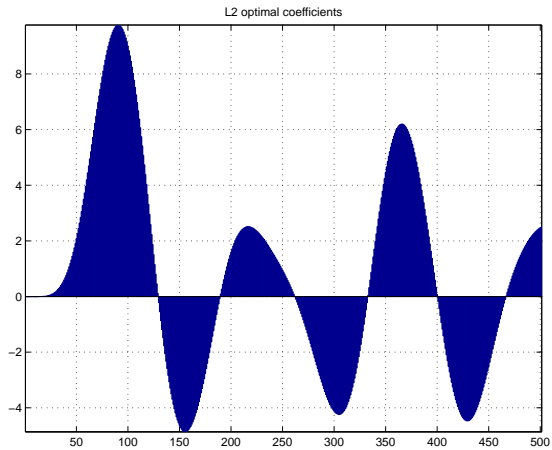}
\caption{Coefficients of $L^2$ spline}
\label{fig:L2coef}
\end{figure}

Fig.~\ref{fig:L1result} shows the resulting fitted curve $y(t)$ 
computed with the $L^1$-optimal control $u(t)$.
We can see that the data are considerably disturbed by Laplacian noise,
but the reconstructed curve well fits the original curve.
To see the sparsity property of the $L^1$-optimal coefficients,
we plot the value of the coefficients in Fig.~\ref{fig:L1coef}.
As shown in this figure,
the $L^1$-optimal coefficients are quite sparse.
In fact, the number of coefficients whose absolute values
are greater than $0.001$ is just $5$ out of $501$ coefficients.
On the other hand, we show the $L^2$-optimal coefficients
with $\lambda=0.0001$, see equation \eqref{eq:J2},
in Fig.~\ref{fig:L2coef}.
This figure indicates that the coefficients are not sparse at all
and the $L^2$ spline requires almost all the base functions
to represent the fitted curve.
\begin{figure}[t]
\centering
\includegraphics[width=\linewidth]{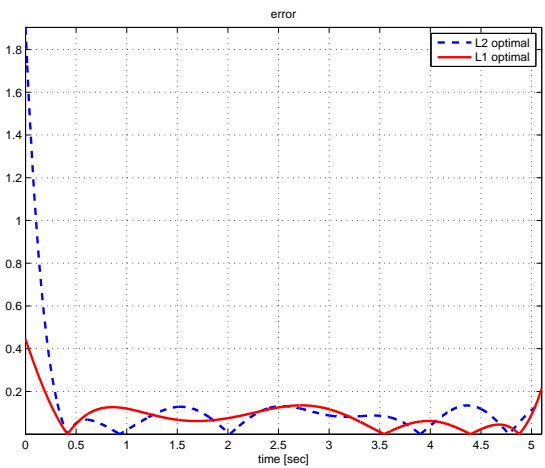}
\caption{Error between original curve $y(t)$ and fitted curve by $L^1$ spline (solid line) and $L^2$ spline (dashed line)}
\label{fig:error}
\end{figure}
Note that the reconstructed curve by the $L^2$ spline also well fits the original curve
as shown in Fig.~\ref{fig:error},
which shows the error between the original curve and the fitted curves.
This figure shows that the $L^2$ spline is almost comparable with the $L^1$ splines%
\footnote{
Another example in \cite{NagMar13WSC}
shows that an $L^1$ spline outperforms
an $L^2$ spline in view of outlier rejection.
}.

In summary, we can say by the simulation that
the proposed $L^1$ control theoretic smoothing spline can effectively reduce the effect of noise in data
and also give sufficiently sparse representation for the fitted curve.

\section{Conclusion}
\label{sec:conclusion}
In this paper, we have proposed the $L^1$ control theoretic smoothing splines
for noise reduction and sparse representation.
The design is formulated as coefficient optimization with
an $L^1$ regularized term and an $L^1$ or $L^2$ empirical risk term,
which can be efficiently solved by numerical computation softwares.
A numerical example has been shown to illustrate the effectiveness
of the proposed $L^1$ spline.

Future work may include extension to constrained splines
as proposed in \cite{NagMar13}, and 
extension to sparse feedback control
as discussed in \cite{NagQueOst14,NagQueNes13}.


%



\section*{Acknowledgment}

This research is supported in part by the JSPS Grant-in-Aid for Scientific Research (C) No.~24560543,
MEXT Grant-in-Aid for Scientific  Research on Innovative Areas
No.~26120521,
and an Okawa Foundation Research Grant.

\ifCLASSOPTIONcaptionsoff
  \newpage
\fi




%

\end{document}